# Geometrical Variation Analysis of an Electrothermally Driven Polysilicon Micoactuator

M. Shamshirsaz*, M. Maroufi*, M.B. Asgari**

*New Technologies Research Centre, Mechatronics Group, Amirkabir University of Technology* (Tehran Polytechnic)
**Energy Engineering Department, Power and water University of Technology, Tehran
*424 Hafez Ave., Postal Code: 1591634311, Tehran-Iran, shamshir@aut.ac.ir, Tel/Fax: +98-21-66402044

*Abstract*-The analytical models that predict thermal and mechanical responses of microactuator have been developed. These models are based on electro thermal and thermo mechanical analysis of the microbeam. Also, Finite Element Analysis (FEA) is used to evaluate microactuator tip deflection. Analytical and Finite Element results are compared with experimental results in literature and show good agreement in low input voltages. A dimensional variation of beam lengths, beam lengths ratios and gap are introduced in analytical and FEA models to explore microactuator performance.

## I. INTRODUCTION

Electrothermal microactuator is generally composed of two suspended beams (arm) joined at the free end. This device generates deflection through asymmetric heating of the hot and cold Polysilicon arms with different cross-section or different length. The cold arm and hot arm are usually made of Polysilicon. When current pass through the microactuator, the higher heat generation in the longer hot arm causes it to heat and expand more than the cold arm. Therefore, this differential expansion forces the tip of device to rotate. The fabrication process is a combination of surface and bulk micromachining technique. The performance of these micro actuators is related to the best rise of temperature to obtain a desired deflection using minimal power.

Many efforts have been carried out to explore the characterization and behaviour of electro-thermally driven microactuators [1-5]. The motivation of all of these efforts is to increase actuator performance by structure optimization. C. Pan and W. Hsu [1] presented a microactuator based on different lengths but the same cross sections for cold arm and hot arm. They investigated length beam effect on the actuator tip deflections by FEA.

In this paper, the influence of geometrical variations on the actuator tip deflections has been explored using analytical and FEA. Different models have been generated and analyzed considering variation of beam lengths, beam lengths ratio and air gaps. An electrothermal model of Polysilicon thermal microactuator similar to Pan's actuator architecture has been developed (fig. 1). To verify the validity of model, analytical and simulation results are compared with experimental results, for similar geometries and input data, presented by C. Pan and W. Hsu. There is a good agreement between these results in low input voltages (about 8 volts). As it was been reported previously by M. Shamshirsaz and M. Gheisarieha [6], the deviation of theoretical results from experimental results in high input voltages is principally due to temperature dependency of Polysilicon properties and particularly Polysilicon thermal expansion coefficient.

## II. FINITE ELEMENT MODEL

A three dimensional analysis of the electrothermally excited microactuators is developed using finite element analysis simulation program. A non-linear analysis is treated considering coupled-field multiphysics analysis. The geometric dimensions and material properties of the microactuator are presented in table 1.

TABLE I

DATA FOR FINITE ELEMENT ANALYSIS

| *Mechanical Properties:* | | *Electric loadings:* | |
|---|---|---|---|
| Young's modulus of polySi | $158 \times 10^9$ Pa | Input voltage | 8 V |
| Poisson's ratio | 0.066 | Resistivity | $5 \times 10^{-4}$ $\Omega$m |
| Density | 2320 Kg m$^{-3}$ | | |
| *Thermal properties:* | | *Geometric dimensions* | |
| Conductivity coefficient | 41 W m$^{-1}$ °C$^{-1}$ | Gap between beams | $5 \leq g \leq 10$ µm |
| Expansion coefficient | $2.7 \times 10^{-6}$ °C$^{-1}$ | Length of long beam | $L_1$=500, 600, 750 µm |
| Specific coefficient | 700 J kg$^{-1}$ °C$^{-1}$ | Length of short beam | $0.1 L_1 \leq L_2 \leq 0.8 L_1$ |
| Convection coefficient | 50 W m$^{-2}$ °C$^{-1}$ | Width of beams | W= 2.8 µm |
| Ambient temperature | 20 °C | Thickness of beams | t=2 µm |
| | | Length of extension beam | L=40 µm |
| | | Contact pads | 200µm×200µm |

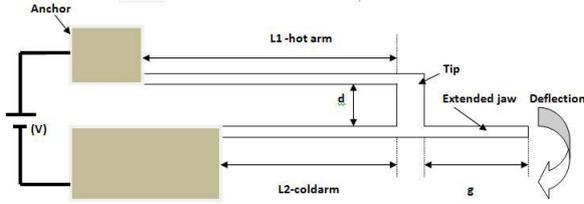

Fig. 1: Electro-thermally driven microactuator designed by Pan and Hsu [1].

### III. GEOMETRICAL VARIATIONS

Microactuator models with three different long beam lengths (L1=500, 600, 750 μm) are developed using analytical and FEA models. Variation of beam length ratios are applied to each model considering long beam length constant and changing short beam length from 0.1 $L_1$ to 0.8$L_1$. For three different long beam lengths model, different models with variable air gap from 5 to 10 μm are generated and analyzed [tab. 1].

### IV. ANALYTICAL MODEL

Pan's actuator architecture is presented in fig. 1. Usually an extended jaw is connected at the end of the microactuator to use it as a microgripper. Analytical analysis of the microactuator consists of two parts; electro-thermal model and thermo-mechanical model that will be discussed in below.

*IV-1 Electro-thermal model*

In the thermal analysis, heat transfer in perpendicular direction of arm axis is ignored because of small cross section comparing to arm lengths.

Heat transfer analysis is investigated considering heat conduction and heat convection. Conduction takes places in the juncture of microactuator to the anchors. Heat dissipations through air also have an important role in thermal analysis. Heat dissipation through radiation is neglected [7].

The coordinate systems for thermal analysis are shown in fig. 2. Under steady state condition the rate of heat generation and losses for a longitudinal element must be equal:

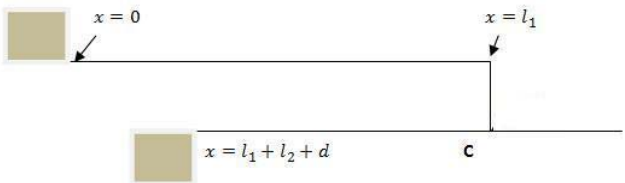

Figure 2: Coordinate systems for thermal analysis of microactuator.

$$-k_p wh \left[\frac{dT}{dx}\right]_x + j^2 \rho wh = -k_p wh \left[\frac{dT}{dx}\right]_{x+dx} + 2(h+w)\beta(T-T_s) \quad (1)$$

Where
dx: Element length
w: Element width
h: Element thickness
β: Convection heat transfer coefficient
$T_s$: Ambient temperature
$k_p$: Thermal conductivity
$j$: Current density
$\rho$: Electrical resistivity of Polysilicon.

Considering temperature distributions on the arms, the linear thermal elongations of hot and cold arms are:

$$\Delta l_h = \alpha \int_0^{l_1} (T_h(x) - T_s) dx \quad (2)$$
$$\Delta l_c = \alpha \int_0^{l_1} (T_c(x) - T_s) dx \quad (3)$$

By solving the electro thermal equations, $\Delta l_h$ and $\Delta l_c$ will be introduced in the thermo-mechanical deflection equations.

*IV-2 Thermo-mechanical model*

The structure of thermal actuator is similar to a plane rigid frame. To solve this statically indeterminate structure, the consistent deformation method is used [8].

By introducing flexibility coefficient $f_{ij}$, compatibility equations can be established and solved simultaneously.

$$\begin{bmatrix} f_{11} & f_{12} & f_{13} \\ f_{21} & f_{22} & f_{23} \\ f_{31} & f_{32} & f_{33} \end{bmatrix} \begin{bmatrix} X_1 \\ X_2 \\ X_3 \end{bmatrix} = \begin{bmatrix} \Delta l_h - \Delta l_c \\ 0 \\ 0 \end{bmatrix} \quad (4)$$

Where $\Delta l_h - \Delta l_c$ are the thermal loads.

Flexibility coefficients can be calculated by Maxwell law due to abundant forces [8]. Calculating redundant forces and moment, moments $M$ along the arm and the tip section can be obtained.
Actuator deflection is calculated applying a virtual unit force and unit moment at point C. Applying consistent deformation method, the virtual moments $M_{v1}, M_{v2}$ along the beam can be obtained. $M_{v1}$ and $M_{v2}$ represent the virtual moment due to unit force and unit moment respectively. According to the virtual work method for the deflection, so the microactuator angle $\theta$ and deflection $u$ can be determined by:

$$u = \sum \int \frac{M M_{v1}}{EI} dx \quad (5)$$
$$\theta = \sum \int \frac{M M_{v2}}{EI} dx \quad (6)$$

The extended jaw's tip deflection $d_{tip}$ for small $\theta$ angles is:
$$d_{tip} = u + g(\theta) \quad (7)$$

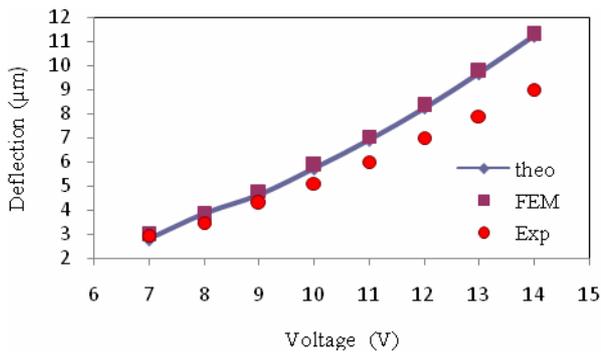

Figure3: Deflection versus input voltages: Theoretical and Finite Element results comparing with expeimental results obtained by Pan and Hsu [1].

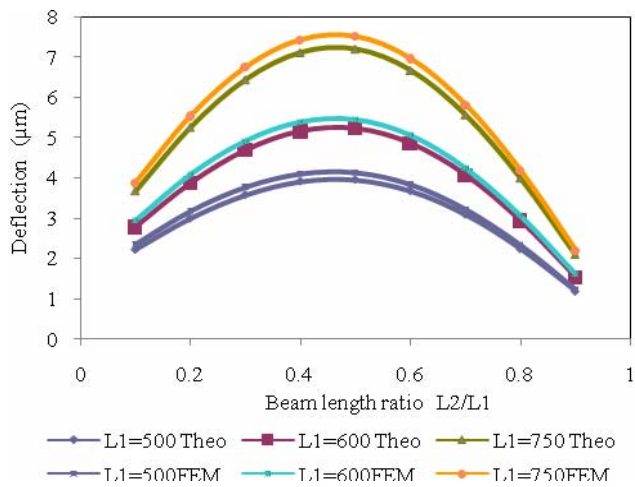

Figure 4: Deflection versus beam length ratio: Theoretical and Finite Element results for an input voltage of 8 volts.

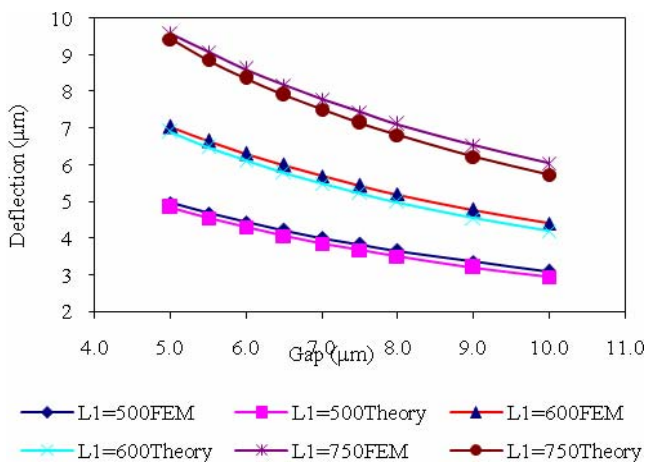

Figure 5: Deflection versus gap: Theoretical and Finite Element results for an input voltage of 8 volts.

V. RESULTS AND CONCLUSIONS

The results show a good agreement between analytical, FEA results and experimental results obtained by Pan and Hsu [1] (fig. 3). By introducing geometrical variations, the results indicate that deflection increases with hot arm length ($L_1$) as it was expected. Analytical and FEA results demonstrate that the microactuator deflection becomes maximum for an optimal beam length ratio value near $L_2/L_1=0.46$ regardless of hot beam length (fig. 4). Pan and Hsu [1] reported an optimal value near $L_2/L_1=0.4$ and 0.5 with lengths of the long beam being 750 and 500 µm, respectively using only FEA. Tip deflection increases with decreasing air gap (fig. 5). For the microactuators with longer hot arm, the tip deflection is more affected by the beam length ratio and air gap variation.

Finally, the results indicate that for current structure, the tip deflection is highly sensitive to geometrical variations. For the longest hot arm in this study, respecting an optimal value of 0.46 for $L_2/L_1$ in microactuator design, an increase of 100% for tip deflection can be achieved. In low input voltages, the variation of geometrical dimensions and air gap influences, principally, heat conduction, heat dissipation by convection and microactuator stiffness.